\documentclass[aps,prl,showkeys,twocolumn,superscriptaddress,groupedaddress]{revtex4}  

\usepackage{graphicx}  
\usepackage{dcolumn}   
\usepackage{bm}        
\usepackage{amssymb}   
\usepackage{latexsym,bm}
\usepackage[tbtags]{amsmath}

\begin{document}

\widetext


\title{Classical and Quantum Kepler's Third Law of N-Body System}

\author{Bohua Sun}

\affiliation{Institute of Mechanics and Technology \& College of Civil Engineering, Xi'an University of Architecture and Technology, Xi'an 710055, China\\sunbohua@xauat.edu.cn}%




\begin{abstract}
\small
Inspired by amazing result obtained by Semay \cite{semay-1}, this study revisits generalised Kepler's third law of an n-body system from the perspective of dimension analysis. To be compatible with Semay's quantum n-body result, this letter reports a conjecture which had not be included in author's early publication \cite{sun2018} but formulated in the author's research memo. The new conjecture for quantum N-body system is proposed as follows: $T_q|E_q|^{3/2}=\frac{\pi}{\sqrt{2}} G\left[\frac{\left(\sum_{i=1}^N\sum_{j=i+1}^Nm_im_j\right)^3}{\sum_{k=1}^N m_k}\right]^{1/2}$. This formulae is, of course, consistent with the Kepler's third law of 2-body system, and exact same as Semay's quantum result for identical bodies.
\end{abstract}

\pacs{45.50.Jf, 05.45.-a, 95.10.Ce}

\keywords{ Kepler's third law; n-body system; periodic orbits; dimensional analysis, classical and quantum mechanics}

\maketitle

The study of the motion between the two bodies under Newtonian gravitational field was solved by Kepler (1609) and Newton (1687) early in the 17th century. For the elliptic periodic orbit of 2-body system, Kepler's third law of the two-body system is given by $T|E|^{3/2}=\frac{\pi}{\sqrt{2}} Gm_1m_2\sqrt{\frac{m_1m_2}{m_1+m_2}}$, where the gravitation constant, $G=6.673\times 10^{-11} m^3kg^{-1}s^{-2}$, the orbit period, $T$, the total energy of the 2-body system, $|E|$, and point masses $m_1$ and $m_2$.
\begin{figure}[h]
\centerline{\includegraphics[scale=0.8]{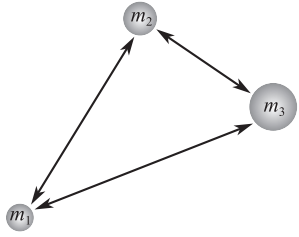}}
\caption{\label{fig-1} 3-body system}
\end{figure}

\begin{figure}[h]
\centerline{\includegraphics[scale=0.8]{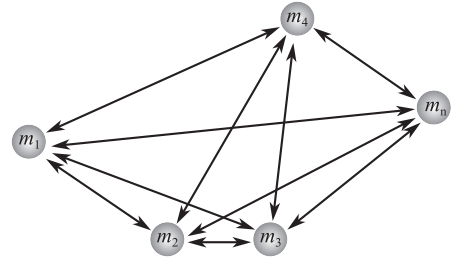}}
\caption{\label{fig-2} N-body system}
\end{figure}

Bohua Sun proposed a conjecture on generalization of Kepler's third law from dimensional analysis \cite{sun2018}. For 3-body system (Fig.\ref{fig-1}), the Sun-Kepler's third law is given by
\begin{equation}\label{p0}
T_3|E_3|^{3/2}=\frac{\pi}{\sqrt{2}} G\left[\frac{(m_1m_2)^3+(m_2m_3)^3+(m_3m_1)^3}{m_1+m_2+m_3}\right]^{1/2},
 \end{equation}
and for N-body system (Fig.\ref{fig-2}) , the Sun-Kepler third law is given by
\begin{equation}\label{p3}
T_N|E_N|^{3/2}=\frac{\pi}{\sqrt{2}} G\left[\frac{\sum_{j=1}^N\sum_{i=j+1}^N(m_jm_i)^3}{\sum_{k=1}^N m_k}\right]^{1/2},
 \end{equation}
where $T_N$ is the period of the orbit and $E_N$ is the total energy (kinetic and potential) of the orbit of n-bodies system.

The Sun's conjecture for classical n-body system in Eq.(\ref{p3}) gives results in good agreement with computations based a great number of periodic planer collisionless orbits for N=3 of \cite{liao-1,liao-2}. Since the publication of \cite{sun2018}, it has received good attentions from \cite{she,zhao,nsr,semay-1}.

Very recently C. Semay \cite{semay-1} reported an amazing result on the quantum supporting of the Sun's conjecture \cite{sun2018}. Semay \cite{semay-1} formulated quantum Kepler's third law for identical bodies as follows
\begin{equation}\label{p1}
  T_q|E_q|^{3/2}=\frac{\pi}{4}Gm^{5/2}N(N-1)^{3/2},
\end{equation}
where the quantum period $T_q$ and quantum energy $E_q$ of N self-gravitating particles with a mass $m$ in D-dimensional space \cite{semay-1,semay-2,semay-3}. For further discussion, we call Eq.(\ref{p1}) as Semay-Kepler's third law.

Semay \cite{semay-1} pointed out the Semay-Kepler's third law in Eq.(\ref{p1}) is similar to the following
\begin{equation}\label{p2}
  T_N|E_N|^{3/2}=\frac{\pi}{2}Gm^{5/2}(N-1)^{1/2},
\end{equation}
that is the result reduced from Eq.(\ref{p3}) for identical bodies where $m_1=m_2=\cdots=m$.

However, the Eqs.(\ref{p1}) and (\ref{p2}) are not exactly the same while with different factors, one has $N(N-1)^{3/2}$ and another has $(N-1)^{1/2}$. The factor difference may implies that the classical Sun-Kepler's third law in Eq.(\ref{p3}) might not be suitable to the quantum n-body system. If this thinking is reasonable, the question will be how to formulate Kepler's third law for the quantum n-body system, which should give the Semay-Kepler's third law in Eq.(\ref{p1}) without the factor difference.

The author went through his research memorandum and seen that, from dimensional analysis and symmetry of mass product, two expressions for n-body system were initially proposed, the first one is Eq.(\ref{p3}) and second one is following
\begin{equation}\label{sun-1}
  T_Q|E_Q|^{3/2}=\frac{\pi}{\sqrt{2}} G\left[\frac{\left(\sum_{j=1}^N\sum_{i=j+1}^Nm_jm_i\right)^3}{\sum_{k=1}^N m_k}\right]^{1/2}.
\end{equation}
For 3-body system, Eq.(\ref{sun-1}) gives
\begin{equation}\label{sun-2}
  T_Q|E_Q|^{3/2}=\frac{\pi}{\sqrt{2}} G\left[\frac{(m_1m_2+m_2m_3+m_3m_1)^3}{m_1+m_2+m_3}\right]^{1/2}.
\end{equation}
Unfortunately, Eq.(\ref{sun-1}) does not give results in good agreement with computations of \cite{liao-1,liao-2}, while Eq.(\ref{p3}) is good agree with \cite{liao-2}. To be compatible with numerical results of \cite{liao-1,liao-2}, only Eq.(\ref{p3}) was reported in the publication \cite{sun2018}.

From dimensional perspectives, both Eq.(\ref{p3}) and Eq.(\ref{sun-1}) are valid invariants. They might be valid for different situations of the n-body system. From Semay's formula on identical bodies in Eq.(\ref{p1}), it would be a natural attempts to apply Eq.(\ref{sun-1}) to the case of identical bodies.

For identical bodies system $m_1=m_2=\cdots=m$, Eq.(\ref{sun-1}) is reduced to
\begin{equation}\label{p5}
\begin{split}
  T_Q|E_Q|^{3/2}&=\frac{\pi}{\sqrt{2}}G\left[\frac{\left(\sum_{j=1}^N\sum_{i=j+1}^Nm m\right)^3}{\sum_{k=1}^N m}\right]^{1/2},
  \end{split}
\end{equation}
noting $\sum_{j=1}^N\sum_{i=j+1}^Nm m= \sum_{j=1}^N\sum_{i=j+1}^N m^2=\frac{N(N-1)}{2}m^2$ and $\sum_{k=1}^N m=Nm$, hence
\begin{equation}\label{pp}
\begin{split}
  T_Q|E_Q|^{3/2}&=\frac{\pi}{\sqrt{2}}G\left[\frac{\left(\frac{N(N-1)}{2}m^2\right)^3}{N m}\right]^{1/2}\\
  &=\frac{\pi}{\sqrt{2}}G\left[\frac{N^3(N-1)^3m^6}{8 N m}\right]^{1/2}\\
  &=\frac{\pi}{4}Gm^{5/2}N(N-1)^{3/2}.
  \end{split}
\end{equation}
It is surprise to see that the Eq.(\ref{sun-1}) predicts exact same result as Semay-Kepler's third law in Eq.(\ref{p1}) for quantum identical bodies system.

Therefore, analogue to classical generalized Kepler's third law in Eq.(\ref{sun-1}), it is reasonable to generalize the Semay-Kepler's third law to a general quantum n-body system as follows
\begin{equation}\label{quantum}
T_q|E_q|^{3/2}=\frac{\pi}{\sqrt{2}} G\left[\frac{\left(\sum_{j=1}^N\sum_{i=j+1}^Nm_jm_i\right)^3}{\sum_{k=1}^N m_k}\right]^{1/2}.
 \end{equation}
The coincidence of classical and quantum results for a generalization of the Kepler¡¯s third law for n-body systems could be more than a simple happy coincidence as Semay stated in \cite{semay-1}, which might implies that Eq.(\ref{quantum}) is a valid conjecture of generalized quantum Kepler's third law for self gravitating particles with unequal mass.

Comparing the above relation with Eq.(5) of Ref.\cite{semay-1,semay-2}, the quantum energy $E_q$ of $N$ self gravitating particles with unequal masses can be predicted as follows
\begin{equation}\label{mass}
  E_q \sim-\frac{1}{2}\frac{G^2}{Q^2\hbar^2} \frac{\left(\sum_{j=1}^N\sum_{i=j+1}^Nm_jm_i\right)^3}{\sum_{k=1}^N m_k},
\end{equation}
where $Q$ is a global quantum number to be determined for unequal bodies. Semay obtained the global quantum number for identical bodies \cite{semay-2}.

In summary, this study indicates that the Kepler's third law will take different forms, namely, Eq.(\ref{p3}) is for classical n-body system and Eq.(\ref{sun-1}) is for quantum n-body system. The reduced expression for identical bodies from Eq.(\ref{sun-1}) fully supports the quantum result obtained by Semay in Eq.(\ref{p1}) \cite{semay-1}.

Although the mechanism behind the happy coincidence is still not clear, this study still brings some new information that could shed some light on this problem. Therefore it is worth to study further.


\end{document}